\documentclass[pre,aps,twocolumn,floatfix,showpacs,showkeys]{revtex4}
\usepackage[dvips]{epsfig}
\usepackage{endnotes}
\usepackage{amsmath}
\usepackage{amssymb}
\usepackage{subfigure}
\usepackage{longtable}

\newcommand{\beq}{\begin{equation}}
\newcommand{\eeq}{\end{equation}}
\newcommand{\bR}{\boldsymbol{R}}
\newcommand{\bta}{\begin{tabular}}
\newcommand{\enta}{\end{tabular}}

\begin{document}

%
%

\title{Structure of plastically compacting granular packings}

%
%

\author{Lina \surname{Uri}}
\email{l.l.uri@fys.uio.no}
\affiliation{Physics of Geological Processes, University of Oslo, P.O.box 1048
Blindern, N-0316 Oslo, Norway.}

\author{Thomas Walmann}
\affiliation{Physics of Geological Processes, University of Oslo, P.O.box 1048
Blindern, N-0316 Oslo, Norway.}

\author{Luc Alberts}
\affiliation{Department of Geotechnology, Delft University
of Technology, Mijnbouwstraat 120, 2628 RX Delft, The Netherlands}

\author{Dag Kristian \surname{Dysthe}}
\affiliation{Physics of Geological Processes, University of Oslo, P.O.box 1048
Blindern, N-0316 Oslo, Norway.}

\author{Jens Feder}
\affiliation{Physics of Geological Processes, University of Oslo, P.O.box 1048
Blindern, N-0316 Oslo, Norway.}

\date{\today}
%
%

\begin{abstract}
The developing structure in systems of compacting ductile grains
were studied experimentally in two and three dimensions. 
In both dimensions, the peaks of the radial distribution function 
were reduced, broadened, and shifted compared with those observed in hard 
disk- and sphere systems.
The geometrical three--grain configurations contributing to the 
second peak in the radial distribution function showed few but 
interesting differences between the initial and final stages of 
the two dimensional compaction.
The evolution of the average coordination number as function of 
packing fraction is compared with other experimental and numerical 
results from the literature.
We conclude that compaction history is important 
for the evolution of the structure of compacting granular systems.

\end{abstract}


\maketitle
%
%

\section{\label{sec:intro}Introduction}
The structure of mono-disperse granular media is known to be very sensitive 
to the shape of the grains \cite{pap:Donev04}, grain interactions
such as cohesion \cite{pap:Xu04,pap:Forsyth01}, and assembling procedure
\cite{pap:Vanel99}. 
Packing fractions as low as $c=0.125$ have been found experimentally 
in 3D-systems with high grain--grain attraction \cite{pap:Bloomquist40},
whereas dense systems of spherical grains can be packed to $c>0.65$ 
\cite{pap:Nowak97} by a carefully selected tapping procedure.

Historically, the structure of dense granular media was studied during the 
1960s and 70s as a model of fluids and amorphous 
materials \cite{pap:Finney70,pap:BernalMason60}.
Since then, the complex properties of granular structures have been studied
in dense elastic packings with much focus on compaction 
dynamics \cite{pap:Knight95,pap:Ribiere05,pap:Richard03,pap:Nowak97}.
The densification of granular packings has a variety of applications, 
and in particular, it takes place in Nature
during the slow compaction of sediments \cite{pap:Weller59} and the fast 
event of a landslide. 
Granular compaction is also commonly studied in relation to 
pharmaceutical powders and metal industry \cite{pap:Fischmeister78}.
In the compaction of sediments and powders, only the initial stages 
can be modeled by the compaction of hard elastic grain ensembles.
When the geometrical structure is jammed \cite{pap:Torquato00} 
(i.e., no grain can be geometrically translated while all others remain 
fixed), at packing fractions $c\sim0.64$, further compaction can only 
occur by grain deformation \cite{pap:Weller59}.
For the compaction of sediments and the hot isostatic pressing of 
metal powders, grains deform plastically, thus deformed regions are 
relaxed during the compaction. 
 
The development of a dense structure in a ductile granular ensemble will 
depend on grain properties such as friction and cohesion, as well as the 
deformation procedure.
In such ductile packings, the densest possible structure 
will depend on how pore space can be reduced at high packing fractions.
At such packing fractions, unless the granular ensemble is compacted in 
vacuum, gas or fluid occupying the pores must be transported out or 
absorbed into the matrix for further compaction to occur.
An example of such a densification experiment is the experiment by 
Stephen Hales in the early 1700s \cite{book:Hales1727}, in which he 
filled a glass with dried peas and water.
The peas absorbed water and eventually filled the container as an ensemble
of polyhedra. 
Very little is known about the structure of plastically deforming 
grain ensembles, although they are extremely important for a range 
of industrially and naturally occurring compaction processes.   

Here, we present a series of experiments on ensembles of compacting
ductile grains in two and three dimensions. 
Positions and coordination numbers have been measured, and 
structural
measures such as the radial distribution function and distribution of 
coordination numbers were obtained. 
In particular, we were able to study the local geometry that dominates
the peaks in the radial distribution function, and the evolution of 
this geometry with packing fraction in two dimensions.
The results are compared to similar findings from the literature of both
hard and ductile experimental grain ensembles, and to a numerical 
compaction model (Rampage),
developed to simulate the initial compaction of sediments 
\cite{thesis:Alberts05}.

The following two sections (\ref{sec:2D},\ref{sec:3D}) presents the 
two and three dimensional systems.
The discussion then follows in section \ref{sec:discussion}.
 
\section{\label{sec:2D}The two-dimensional system}
The two-dimensional (2D) system consisted of 1100 oriented cylinders, 
which were compacted at constant rate.
From image analysis, the positions, coordination numbers, and 
packing fractions were found at regular intervals during the compaction.
\subsection{\label{sec:exp2d}Experiment}
The  2D-setup consisted of ductile cylindrical grains, which were stacked 
in a Hele-Shaw cell, and uniaxially compressed by a piston at a constant 
rate $v=1$ mm/h.
The grains were made of spaghetti, which had been boiled in 
water for 14.5 minutes in order to render them soft and deformable.
The grains' resistance to deformation could be controlled by the boiling
time.
After boiling, the spaghettis were kept in a clear olive oil to avoid 
dehydration, then cut into cylinders of length 10 mm, and carefully stacked 
in the container (Hele-Shaw cell).
Two sizes of grains were used, but in different sections of the model.
For simplicity, only the lower section of nearly 600 mono-disperse
grains was used in this analysis. 
These grains had (initial) diameters of $d=(2.8\pm 0.1)$ mm. 

\begin{figure}
\epsfig{file=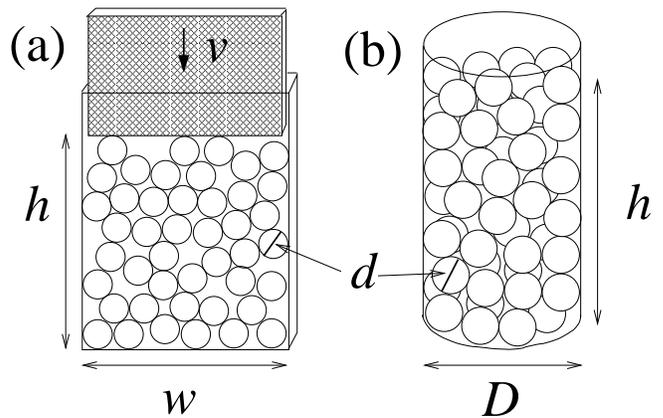,width=8.6cm}
\caption{Schematic illustrations of the experimental setups. $h$ is the
filling height, and $d$ is the grain diameter in both 2D and 3D. 
(a) The quasi two-dimensional arrangement. $w$ is the width of the 
container, and the piston is driven at constant velocity $v$.
The grains were cylinders made from spaghetti, and stacked such that
their length was in the depth of the container. 
When viewed from the front of the container 
only the circular crossection of the grains were seen, of diameter $d$.
(b) The three-dimensional cylindrical ensemble, consisting of spherical 
grains of diameter $d$ made of Playdoh. The inner container 
diameter is $D$.}
\label{fig:setups}
\end{figure}

The width and height of the initial packing was $w=83$ mm and $h=92$ mm, 
corresponding to 29 and 33 grain diameters, respectively.
The initial height of the 600 lower grains was 55 mm, or 20 grain 
diameters.
During the stacking, olive oil was constantly added to the container 
so that the grains were immersed in oil at all times.
The oil, apart from preventing dehydration of the grains, also lubricated
the walls so that friction did not restrain the compaction.
The depth of the Hele-Shaw cell was 11--12 mm, to ensure a channel at the
back of the stacked grains for the oil to escape through. 
The oil thus did not affect the compaction mechanically. 
Figure \ref{fig:setups}(a) illustrates the setup of the 2D 
experiment, and Fig. \ref{fig:zoom} contains a closeup of a region in 
the first image.
Pictures were taken every 20 minutes, and a total of 47 pictures were
taken as the system compacted at a speed of $v=$ 1 mm/hour from an initial 
packing fraction of $c=0.89$ to the final $c=0.99$.
The camera was an AstroCam Capella, LSR Life Science Resources, UK, 
with 3000$\times$2000 pixels, 14 bit resolution.
The spatial resolution was 23 pixels per mm. 

\subsection{\label{sec:analysis2} Analysis (2D)}
\begin{figure}
\epsfig{file=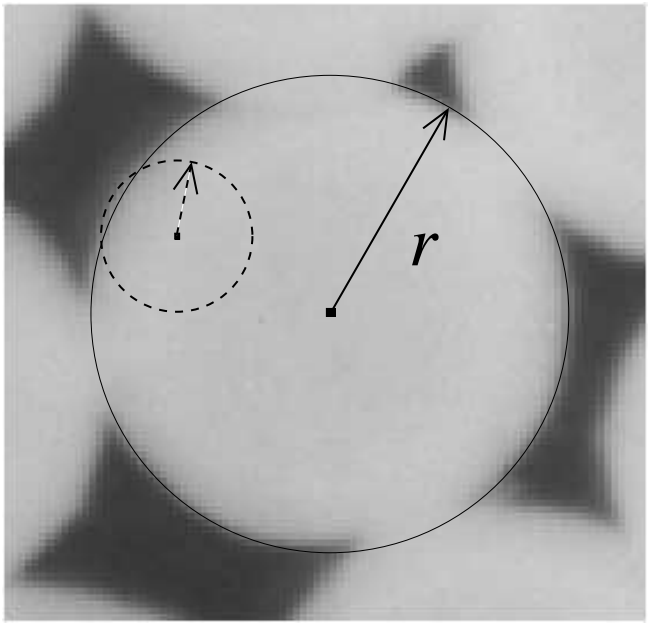,width=8.6cm}
\caption{Closeup of a grain in the first image.
The procedure of finding the grain center in the first image is illustrated: 
the two circles represent the largest distance from their
center pixels at which a maximum of 5 ``dark'' pixels are found (see text for 
details).}
\label{fig:zoom}
\end{figure}

The geometrical center position was obtained for all grains  
by image analysis.
In the first image, i.e., the image taken immediately after compaction 
started, the center of mass was found by the following 
procedure:
considering circles around a particular pixel, the radius of the 
circle that contained 5 pixels of intensity below a certain threshold 
was taken to be the shortest distance to the pore space from this 
particular pixel.
The pixel within a grain with the {\em largest} distance to the pore space 
was taken to be the center position of the first image, see 
Fig. \ref{fig:zoom} for an illustration.
None of the disks were much deformed at this stage, so looking for pore
space in circular sections accurately determined the center position 
to a resolution of a fraction of a pixel for most grains.
However, due to reflected light from the Hele-Shaw cell, some positions 
were only determined within a few pixels resolution. 
The center of mass positions found in the first image were traced by 
pattern recognition (normalized cross correlation function 
\cite{book:Pratt91}) in all the following images, to a resolution of 1 pixel.

The grains were essentially incompressible. 
During compaction, the average crossectional area $A$ of each grain 
decreased 6\%, thus the grains elongated in the direction perpendicular 
to the image plane.
The development of the crossectional area was found manually for 
ten specific grains in the first and last image.
Of these ten grains, two representative grains were selected, and their areas
were found (also manually) in every second image throughout the experiment.
A best fit was made to the developing average area, as
$A_m=a(1-(m/b)^2)$, where $m$ is image number, $a=3.6\,10^3$ square 
pixels, and $b=195$.
The effective grain diameter $d(m)$ in each image was then found as
$d(m)=\sqrt{4A_m/\pi}$.
\begin{figure}
\epsfig{file=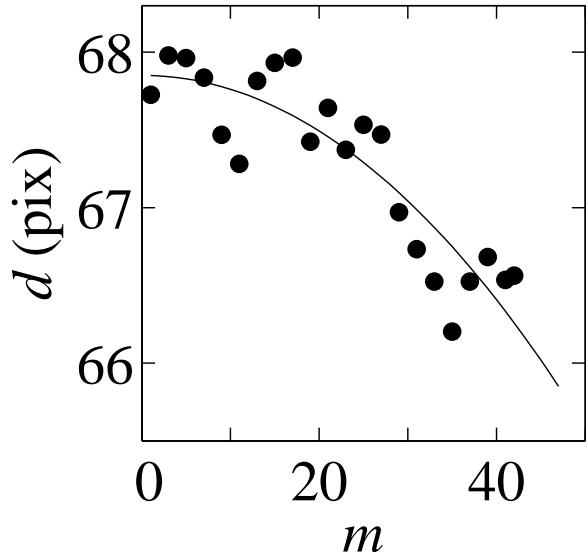,width=8.6cm}
\caption{The effective grain diameter, $d(m)$, (line) as function of 
image number, $m$. 
($\bullet$) Grain diameters,$d=\sqrt{4A/\pi}$, calculated from the 
average of the measured area, $A$, of two grains.}
\label{fig:bestfit}
\end{figure} 
Figure \ref{fig:bestfit} shows the grain diameter, calculated from the 
average area of each image, and the effective diameter $d(m)$ (line) 
as functions of image number $m$.
By using the effective diameter in all the analysis, the decreasing 
crossectional area does not affect the geometry and structural 
evolution during the compaction.
Thus, the packing fractions and radial distribution functions can be 
compared between images independent of the actual grain diameters.

\subsubsection{\label{sec:c2d}Packing fractions}
\begin{figure}
\epsfig{file=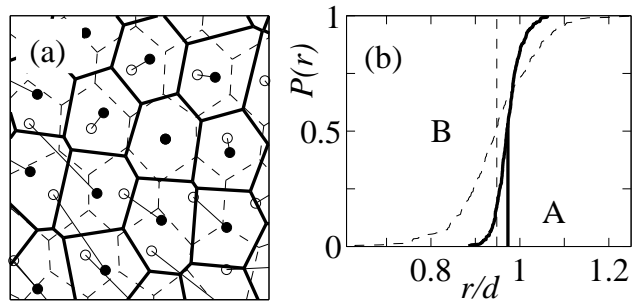,width=8.6cm}
\caption{(a) An example of the Voronoi structure in a region of the first
and last (dashed) images, used in obtaining the local packing fractions $c_i$.
The tessellation of the last image was translated so that the center of one 
cell lie undistorted. The compaction direction is downward.
The cumulative distributions of touching neighbor distances are shown in (b), 
for the first (A) and last (B, dashed) images. The average value of distances 
between touching neighbors is indicated by vertical lines for the first
and last (dashed) image.}
\label{fig:tesselation}
\end{figure}

The packing fraction $c$ of the 2D system was obtained
by Voronoi analysis of the position data from each image.
Only disks at a distance of more than one grain diameter from any boundary 
were used in the calculation to avoid unbounded \cite{man:matlab} Voronoi 
cells. 
The area $A_{vi}$ of the Voronoi cell for each disk center $i$ was found, 
and as each such cell contains one disk, the local packing fraction 
$c_i$ was given by 
\beq
c_i = \pi  d^2/4A_{vi},
\eeq
where $d$ is the average diameter of a grain in the image in question.
The Voronoi tessellation of a region around a certain grain is shown for 
the first and last (dashed) image in Fig. \ref{fig:tesselation}(a), 
with grain centers marked as bullets (first image) and circles (last 
image). 
The motion relative to the central grain is indicated by lines 
between the grain center positions in the first and last image.
Figure \ref{fig:tesselation}(b) shows the cumulative distribution 
$P(r)$ of distances between touching grains for the first (curve A) and
last (curve B) images.
The overall packing fraction based on the Voronoi cell division was
$c=\langle c_i\rangle$.
The 2D compaction spanned packing fractions from $c=0.89$
to $c=0.99$.

\subsubsection{\label{sec:rdf2d}Radial distribution function}
The radial distribution function (RDF) $g(r)$ is defined by 
$g(r)=\rho (r)/ \langle \rho \rangle $, where $ \langle \rho \rangle =N/V$ is 
the average number density of grain centers in a container of area $V$, 
and $\rho (r)$ is the average number density as 
a function of distance $r$ from a grain center. 
The expression for the number density as measured from grain number $i$ is 
\begin{eqnarray}
\rho_i(r)=\frac{N_i(r)}{V_n(r)}\, ,\\
 N_i(r)=\sum_{j\neq i}\delta(r_{ij}=r)\, .
\end{eqnarray}
Here, $V_n(r)=2\pi r$ is the circumference of a circle of radius $r$, 
and $N_i(r)$ is the number of grain centers at distance $r$ from 
grain $i$.
$r_{ij}$ is the distance between grains $i$ and $j$, and $\delta$ is the 
Kronecker delta,
$\delta =1$ for $r_{ij}=r$, and otherwise zero.
The average $\rho_i(r)$ over $N$ grains gives
\beq
\rho(r)=\frac{1}{N}\sum_{i=1}^{N}\rho_i(r)\,,
\eeq
thus
\beq
g(r)=\frac{1}{N\langle\rho\rangle V_n(r)}\sum_{i}N_i(r)\,,
\label{eq:defRDF}
\eeq
where $i$ is summed over all grain centers of the sample.
In a finite system the expression for $g(r)$ should be replaced by
\beq
g(r)=\frac{1}{N\langle\rho\rangle V_n(r,dr)}\sum_{i}N_i(r\in[r-dr/2,r+dr/2])\, 
,
\label{eq:defRDFfinite}
\eeq
where $dr$ is the width of the shell, its size depending on the number 
of grains in the system and/or the uncertainty of the position measurements.
Perturbations to $dr$ should not affect $g(r)$ when the proper value of 
$dr$ is chosen, as  the dependence of $N_i$ and $V_n$ on $dr$ should cancel 
in the expression of $g(r)$.
The expression for the circular shell becomes 
$V_n(r,dr)=\pi [(r+dr/2)^2-(r-dr/2)^2] = 2\pi r\,dr$. 

Noting that the sum in Eq. (\ref{eq:defRDFfinite}) over $N_i$ equals 
twice the number of distances $n(r,dr)$ of lengths $r\in[r-dr/2,r+dr/2]$ 
in the ensemble, $\sum_{i}N_i(r,dr) = 2n(r,dr)$, 
the expression for the RDF of a finite packing becomes 
\beq
g(r)=2n(r,dr)F(r)/(N\langle\rho\rangle^2 V_n(r,dr))\,,
\label{eq:rdf_distFinite}
\eeq
where $F(r)$ is a finite size correction for the boundaries, as discussed 
below.

When Eq. (\ref{eq:rdf_distFinite}) is used without the normalization $F(r)$, 
the boundaries of the ensembles introduce finite size effects to the RDF 
of a small system.
These finite size effects result not only because the structure along the 
boundaries differ from the interior structure, but because the normalization 
over circular shells of areas $2\pi r dr$ includes regions outside of the 
ensemble.
To avoid the latter source of errors, the normalization function $F(r)$ is 
introduced based on the specific rectangular geometry of the ensemble.
$F(r)$ is found as follows:
In an infinite system, a grain in a position $\bR=(x,y)$ would be surrounded 
by a circular shell $2\pi r\, dr$ independent of its position. 
In the finite system of a container, whether all of the circular shell 
lies within the container depends on the radius $r$ and the grain's position 
$\bR$.  
The fraction of circular shells of radius $r$ that partly lie outside 
of the container increases with $r$ and is a measure of the error done 
by disregarding the boundaries. 
Let the area of the container be $V$. 
The integral over $V$ of all circular shell circumferences $2\pi r$ with 
center positions inside the area of the container is $2\pi r V$. 
Only a part of this integral represents circumferences that lie inside 
the container, thus in 
normalizing the RDF of a finite system one should use this fraction instead 
of the $2\pi r V$.
A normalization function $F(r)$ can be defined as the ratio of the 
integrals of circumferences in the infinite case to the 
finite case:
\beq 
F(r)=\frac{\int_{V}2\pi r\, d^2{\bR}}{\int_{V}A(r,\bR)\, d^2\bR} 
\label{eq:F(r)}
\eeq 
$A(r,\bR)$ is the fraction of the circular shell of radius $r$ centered
at $\bR$ that lies inside the container.
For a rectangular container of width $w$ and height $h$ the normalization 
function is:
\beq
\begin{split}
F(r)=\pi wh\bigl[\pi (w-2r)(h-2r)+&(\pi -1)(w+h-4r)\\
&+r(3\pi /2 -2)\big]^{-1}\,.
\label{eq:F(r)2d}
\end{split}
\eeq
This correction procedure was previously used by Mason \cite{pap:Mason68} 
for a different geometry (3D spherical ensemble).

\subsubsection{\label{sec:coord2d}Coordination numbers}
\begin{figure}
\epsfig{file=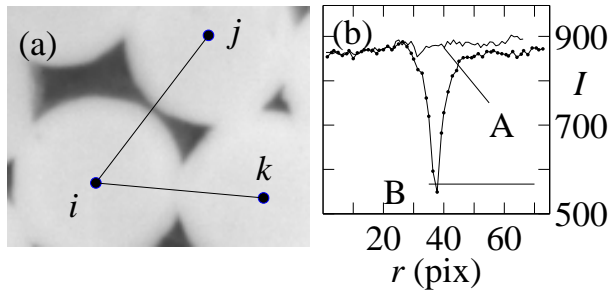,width=8.6cm}
\caption{The procedure of finding touching neighbors was based 
on the intensity values of the direct line between grain centers.
(a) shows a small region of the initial compaction state, where 
$i,\,j$, and $k$ are grain centers. The lines $i$--$j$ and $i$--$k$
are the direct lines between the centers, of which the intensities
are plotted in (b) as curve A and B, respectively.
The dip shown in curve B was typical when grains did not touch.}
\label{fig:2Dcoord}
\end{figure}

The coordination numbers $k$ (number of contacts per grain) were found 
from further image analysis; 
a Delaunay triangulation based on the position data was used to obtain the set 
of nearest neighbors of each grain.
Each Delaunay neighbor connection was then investigated by a thresholding 
procedure to establish whether it was a touching neighbor:
The intensity values of the array of pixels that formed the shortest 
path between grain centers $i$ and $j$ showed a distinct dip whenever
pore space was present between the grains.
Grains were considered not touching if the dip was below 2/3 of the intensity
value of the grain face intensity. 
Figure \ref{fig:2Dcoord} illustrates this procedure.
Grains $i$ and $j$, as seen in Fig. \ref{fig:2Dcoord}(a), are not touching 
as the minimum value of their center-to-center
intensity plot is below $I=850\cdot 2/3=567$, as seen in (b). 
The intensity values was in the range $I\in[100,1000]$, where $I=100$ 
represented black. 
The typical difference in intensity values between open pore space and 
the internal of a grain was $\Delta I=500$.  
The coordination number distribution $P(k)$ was found with this 
procedure for all the images.

\subsection{\label{sec:results2d}Results (2D)}
\begin{figure}
\epsfig{file=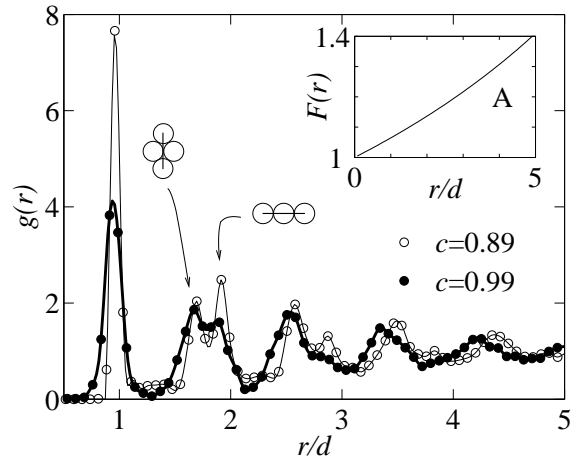,width=8.6cm}
\caption{The radial distribution function, $g(r)$, as a function of 
particle separation $r$ in the ductile 2D experiment, for packing fractions
$c=0.89$ ($\circ$) and $c=0.99$ ($\bullet$, bold line).
The curves connecting the data points are produced by splines.
The normalization function for the 2D setup, $F(r)$, Eq. (\ref{eq:F(r)2d}), 
is shown in the upper right corner (curve A) for the distances $r/d\in[0,5]$.
The typical geometrical configurations that contribute to the peaks at 
$r/d\in[1.695\pm 0.05]$ and $r/d\in[1.915\pm 0.05]$ are illustrated.}
\label{fig:RDF2d}
\end{figure}

The radial distribution functions for the first and last stages of the 
experiment are shown in Fig. \ref{fig:RDF2d}.
The structure at $c=0.89$ shows many of the characteristic peaks of
a triangular packing of disks, except that the peaks are broadened and
shifted to shorter distances than those of a crystalline packing. 
Crystalline correlations occurring in triangular 
lattices of hard disks at $r/d=(1,\sqrt 3,2,\sqrt 7,3)$
are shifted to $r/d=(0.958, 1.695, 1.915, 2.580, 2.875)$.
In the RDF of the more compacted system of $c=0.99$ the peak near $r/d=3$ has
vanished, and the double peak near $\sqrt 3 \simeq 1.7$ has become smoother 
and broader.
Also the first peak ($r/d=1$) is seen to broaden considerably as many neighbor
distances at this stage are less than one effective grain diameter.
The inset of Fig. \ref{fig:RDF2d} shows the finite size correction function 
$F(r)$ for the first image, which is practically identical to the correction 
in the last image.

\begin{figure}
\epsfig{file=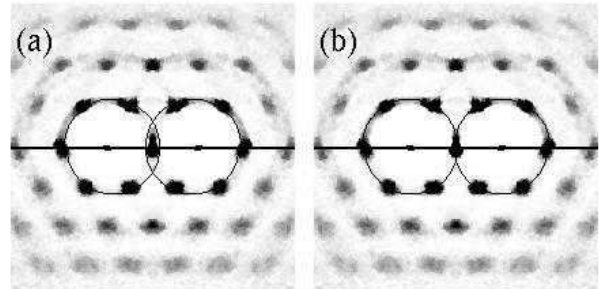,width=8.6cm}
\caption{Visual illustration of the probability of a third 
grain position for two grains at a given distance (a) $r=$1.645--1.745 $d$, and
(b) $r=$1.865--1.965 $d$. The upper part of the plots show 
the probability distribution at $c=0.89$, while the lower give the 
distributions at $c=0.99$.}
\label{fig:geom}
\end{figure}

The geometrical structures associated with the peaks at $r/d=1.695$ and 
$r/d=1.915$ in the RDF are illustrated in Fig. \ref{fig:geom} by a 
conditional three-point correlation function \cite{pap:Torquato02}.
Figure \ref{fig:geom}(a) was generated by identifying all
grain centers at distances $r/d\in [1.695\pm0.05]$ to each other in the 
packing, aligning
each pair of such centers along the horizontal axis, and plotting the 
surrounding grain centers in the corresponding positions.
The two fixed grain positions are shown on the horizontal axis at the 
center of the figure, while the surrounding intensity structure visualizes
the probability of having a third grain in any position relative to the 
two fixed ones.
The intensity was normalized by the number of pairs contributing to the plot.
Figure \ref{fig:geom}(b) is generated in the same manner as (a), but 
for grains at a distance of $r/d\in [1.915\pm 0.05]$ to each other.
The upper half of (a) and (b) shows the probability distribution for the 
initial stage of the compaction ($c=0.89$), while the lower part shows
the distribution at $c=0.99$. 
Black represents a high probability of the third grain, and white 
represents zero probability.
The black circles mark the distance of one diameter from the two fixed
grain positions.
The intensity pattern suggests a regular structure with obvious
symmetries in the packing.
At first glance, the structure might seem hexagonal, but this is  due
to the alignment of the two fixed centers; 
The underlying structure is triangular. 
Although the intensity plot of Fig. \ref{fig:geom}(a) does not prove 
the presence of the four-point configuration 
illustrated in Fig. \ref{fig:RDF2d}, closer inspection of
the configurations contributing to this peak (at $r/d=1.695$) in the RDF
shows that this configuration is dominant for $r/d=1.695$.
Similarly, the configuration of three grains in a row is seen to contribute
most to the peak at $r/d=1.915$ of the RDF, although the center grain 
is often slightly misplaced along the axis.
The lower part of Fig. \ref{fig:geom}(a) contains a small probability of 
a third grain in between the two fixed grains (barely visible in the figure), 
which is not seen in the upper part ($c=0.89$).
This very compressed alignment of three grains in a row emphasize the 
role of ductility of the grains, as this configuration is impossible in
packings of hard disks.

The cumulative distribution $P(r)$ of distances between touching grains is 
shown in Fig. \ref{fig:tesselation}(b).
The average distance between touching grains decreases from the first
to the last image, while the width of the distribution increases.
The touching grains at distances larger than a grain diameter are 
grains aligned perpendicular to the compaction direction, thus the 
ellipticity of grains increases, especially in the last compaction 
stage.

\begin{figure}
\epsfig{file=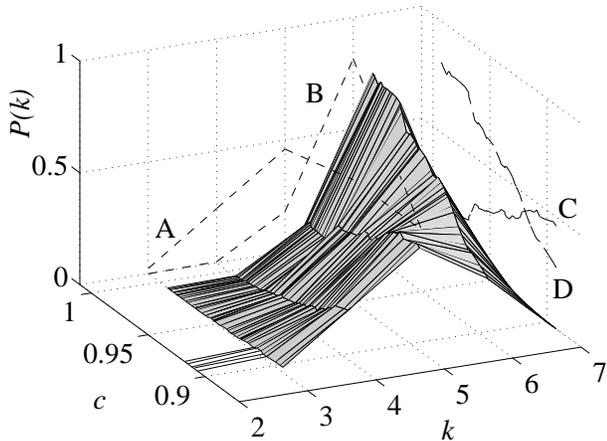,width=8.6cm}
\caption{The distribution $P(k)$ of coordination numbers $k$ as function of 
packing fraction $c$ in two dimensions. Curves A and B are projections of 
$P(k)$ at $c=0.89$ and 0.99, respectively. Curves C, for $P(5)$, and D,
for $P(6)$, are developments of 
$P(5)$ and $P(6)$ with increasing $c$.}
\label{fig:Cdev2d}
\end{figure}

The coordination number distribution is shown in Fig. \ref{fig:Cdev2d} as
a three-dimensional plot of the distribution $P(k)$ as function of 
coordination numbers $k$ and packing fraction $c$.
The distribution is broad at $c=0.89$, and then narrows and shifts to higher 
coordination numbers with increasing packing fraction.
At packing fractions $c=0.906$--$0.918$ a few cases of grains with only two
contacts are observed, but only one or two such grains are present at each
packing fraction.
Curves A and B in Fig. \ref{fig:Cdev2d} are projections of the first and 
last distributions of the experiment, respectively.
Curves C and D are projections of the fractions of five and six coordination
numbers with increasing packing fraction. 

\begin{figure}
\epsfig{file=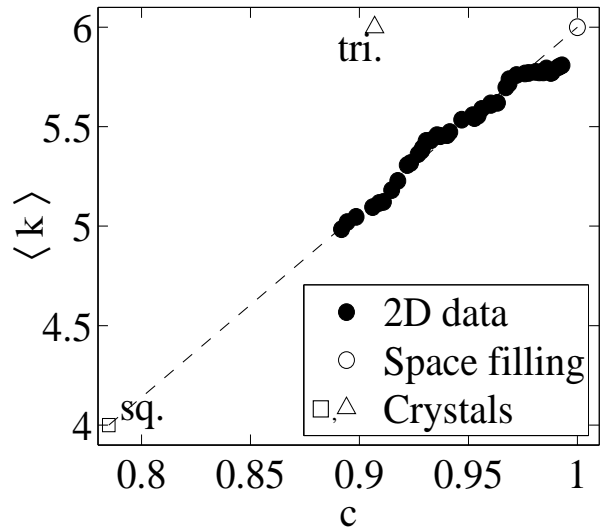,width=8.6cm}
\caption{Mean coordination numbers $\langle k\rangle$ as a function of 
packing fraction $c$ for the two dimensional ductile experiment. 
The circle represents the 2D space filling structure.} 
\label{fig:cvsC2d}
\end{figure}

Figure \ref{fig:cvsC2d} shows the average coordination number 
$\langle k\rangle$ as function of packing fraction in the two dimensional 
experiment. 
The circle at coordinates (1,6) represents the endpoint for 2D systems,
as a space filling structure (Voronoi cells) in 2D must have six neighbors
on average \cite{book:Coxeter69}.
Also shown are the data for two crystalline structures, the square
lattice and the triangular lattice.
We observe that the evolution of the average coordination number in the 
ductile packing closely follows a straight line between the square lattice 
and the space filling Voronoi structure. 

\section{\label{sec:3D}The three-dimensional systems}
The three-dimensional (3D) systems analyzed here are six experimental 
systems of compacting ductile grain ensembles. 
These are compared to Finney's experiment \cite{pap:Finney70} on a random 
close packed ensemble of hard spheres ($c=0.6366$), from which the positions 
were kindly made available to us.
This ensemble contained nearly 8000 grains in a square box, and the positions
of the grains were measured to a precision of 0.2\% of the grain diameter.
A numerical model of a compacting grain ensemble is also presented, giving
possible trajectories for compaction between packing fractions 
$c = (0.50$--0.64). 
\subsection{\label{sec:exp3d}Experiments}
The 3D experimental setup consisted of 
$N$ ductile grains poured into a cylindrical Plexiglas container, 
approximately ten at a time, to a height $h$. 
The set of grains were then left to compact under gravity for 
a time $t$, or were mechanically compacted to a specific 
packing fraction.
A lid covered the top of the cylinder during the experiment to avoid 
dehydration, as the grains contain water and other volatile chemicals.
Figure \ref{fig:setups}(b) illustrates the setup of the 3D 
experiments.
The grains were prepared from Play-Doh
(Hasbro Int.~Inc., UK) to spheres of  diameters $d=(8.8 \pm 0.2)$ mm. 
The material is viscous for shear 
rates in the range ($10^{-3}$--$10^{-6}$)/s, with a viscosity of 
$3\cdot 10^{5}$ Pa\,s, as measured in rotational tests (Paar-Physica 
MCR 300, rheometer).
Except from the first few minutes and the very end of an experiment, the
strain rate was within this regime. 
Strain rates down to $10^{-7}$/s were common after a week's compaction, 
but shear tests could not be performed at such low rates, and the 
material properties for these rates are not known.

Each experimental system was disassembled grain by grain while measuring
positions and/or coordination numbers of the ensemble. 
The position coordinates ($x,y,z$) of the grains' top point were established 
with a mechanical arm (MicroScribe) to an estimated resolution of 
$0.5$ mm (6\% of grain diameter), limited by the difficulty of 
establishing the center of the grain's top surface.
The number of touching neighbors (coordination number) was also 
counted for each grain during the disassemblage.
To ease this procedure, grains of five different colors were used, 
as grains of different colors left marks on each other in contacting areas.
Before a grain was removed, the visibly contacting neighbors were counted.
The grain was then removed, and   
any formerly hidden contacts became visible due to grain deformation and/or 
dis-coloring. 
As a grain was removed, all its contacting grains that remained in 
the ensemble were carefully marked with a felt pen. 
Thus, the coordination number of a grain consisted of the number of 
marks on its upper surface, plus the number of visible and hidden 
contacts established during the removal of the grain itself.   
The number of contacts that each grain had to the cylinder wall 
and/or floor was also recorded.

Six different three dimensional ensembles were studied: 
Two of them with 2200 grains in a cylinder of diameter $D=130$ mm, the 
rest with 150 grains in cylinders of diameter $D=32$ mm. 
The four small systems were prepared simultaneously, and left to 
compact for different amounts of time before each was disassembled
and the positions and coordination numbers of each system measured. 
The resulting packing fractions were $c=$0.64, 0.64, 0.66 and 0.68 
$\pm 0.01$.
The small system size was desirable in order to avoid considerable 
compaction during the disassemblage for the short compaction times. 
One of the big systems was left to compact for 145 hours before it was
disassembled, measuring the positions and coordination numbers of the 
grains.
The packing fraction went from 0.57 (based on filling height) to 0.701 
during the compaction in this ensemble.
Due to a calibration error with the mechanical arm, only the positions of 
787 grains in the mid to lower part of the cylinder were successfully 
measured, while the coordination number was established for 1169 grains. 
The second large system was compacted to a packing fraction of $c=0.75$ by 
gently pushing a piston from the top of the cylinder.
Compacting the system to such high packing fractions would otherwise 
have required months of gravitational compaction.
Only the coordination numbers were measured in this ensemble, for 
839 central grains.

\subsection{\label{sec:finney}The Finney ensemble}
Finney \cite{pap:Finney70} carefully measured the positions 
of nearly 8000 mono-disperse spheres in a box of square crossection.
The spheres were steel ball-bearings of diameters 1/4 inch, 
which positions were measured to $0.2\%$ of the grain diameter.
The central 817 grains were used in this analysis, as 
these grains had a minimum distance of five grain diameters to 
any boundary, thus no corrections to the RDF was necessary to 
$r/d=5$.
Based on this selection, the radial distribution function and 
coordination numbers were found.

\subsection{\label{sec:rampage}Numerical compaction model}
A numerical model (Rampage, \cite{thesis:Alberts05}) was used to simulate 
dry granular media. 
It consisted of randomly placing one thousand  mono-disperse 
spherical grains of diameter $d$ in a container of square cross-section.  
Periodic boundary conditions applied in the horizontal plane. Grains were 
randomly positioned until a packing fraction of $c= 0.5$ was obtained.
No considerations about overlapping particles were done during this initial 
filling procedure. 
The overlaps were then reduced  by 
repositioning grains according to a force balance based on the elastic 
energy of the overlap region, and gravity. 
An overlap of volume $V_o$ resulted in an inter-grain force $f$ of
\beq
f=\frac{YV_o}{2d},
\label{eq:elasticOverlap}
\eeq 
in which $Y$ is Young's modulus of the material.
When gravitationally unstable, a grain would drop or roll until a 
gravitationally stable condition was obtained. 
This compaction process went on until a predefined toleration threshold of
average overlap distance $l$ was reached ($l/d < 0.005$), while the 
ensemble was gravitationally stable.
At this point further compaction was obtained by {\textit {tapping}}, an 
incremental 
($<0.001h$) vertical re-positioning downward of each particle 
(at height $h$), followed by the same procedure of reducing overlaps 
and reaching gravitational equilibrium.
Packing fractions of stable numerical ensembles were in the range
$c=0.55$--$0.64$.
This model was originally developed for the modeling of sediment compaction,
and is described in detail in Ref. \cite{thesis:Alberts05}.
The model reproduces statistically Finney's ensemble at $c=0.636$ 
with respect to the RDF, the coordination number distribution and 
the distribution of contact angles, see Ref. \cite{thesis:Alberts05}.

\subsection{\label{sec:analysis3}Analysis (3D)}

\subsubsection{\label{sec:c3d}Packing fractions}
In three dimensions, the packing fractions were computed from the 
position measurements for all the cases in which these were available, 
which were most of the ductile experiments, the Finney ensemble, and 
the numerically generated ensembles.
Boxes of variable square cross-section $l^2$ and height $h$ equal to 
the height of the packing were centered in the granular
ensemble for calculation of the packing fraction.
For each of these boxes $j$, the packing fraction was 
$c_j(l)=N_jV_g/V_{bj}$, where $N_j$ is the number of grains 
contained in the box, 
$V_g=\pi d^3/6$ is the average volume of a grain, and
$V_{bj}=l^2h$ is the volume of the box.
Grains that partially intruded the box, i.e., when their center position
was less than one grain diameter from the box boundary, also contributed
to the number of grains in the box.
The overall packing fraction $c$ of an ensemble was obtained by averaging over
$n=10$ boxes in distances 0.5--1.5 grain diameters $d$ from the 
boundary: 
\beq
c=\frac{1}{(n+1)}\sum_{j=0}^{n}c_j(l_0+j\Delta l).
\eeq
where $l_0$ is the side of the box in a distance
1.5 grain diameters from any boundary, and $\Delta l = d/n$.
In the numerical model, the total volume of two touching grains was 
smaller than two grain volumes if the grains overlapped.
The overlap volume was not assumed to be transported to the pore 
volume, as this would effectively change the structure of the packing, 
thus the calculation of the packing fraction must account for the 
excess overlap volumes in the packing.
This was done by reducing the sphere radii by the mean overlap 
distance in the packing, thus reducing the volume of the grains
before calculating $c$.

\subsubsection{\label{sec:rdf3d} Radial distribution function}
The radial distribution function in three dimensions is given by 
Eqs. (\ref{eq:defRDFfinite}) and (\ref{eq:rdf_distFinite}), but with $V_n$ as 
the volume of a spherical shell of radius $r$ and width $dr$ in 3D: 
$V_n=4\pi[(r+dr/2)^3-(r-dr/2)^3]/3 = 4\pi r^2\, dr + \pi(dr)^3/3$.

Equation (\ref{eq:defRDFfinite}) was used in the calculation of the RDF
in the Finney ensemble, as an ensemble average was made only over the 
central 817 grain positions to avoid boundary effects. 
These grains were all in a minimum distance of five grain diameters from 
any boundary, thus the RDF is not affected by boundaries up to $r/d=5$. 
In order to calculate the RDFs of the large ductile ensemble ($c=0.70$),
Eq. (\ref{eq:rdf_distFinite}) was used, as the 
statistics of the whole ensemble was needed for a sufficiently detailed
RDF to be found.
A finite size correction function $F(r)$ was introduced for the ductile 
ensemble in the same manner as for the 2D system, Eq. (\ref{eq:F(r)}).
In 3D, $A(r,\bR)$ is the fraction of the spherical shell area (radius $r$, 
center position at $\bR=(x,y)$, height $z$) which is inside the 
cylindrical container.
The resulting correction function is
\beq
F(r)=\frac{4\pi ^2r^2R^2h}{\int_0 ^{\text{min}(h,r)}(\text{I}_1-\text{I}_2)dz}\,,
\label{eq:F(r)PD}
\eeq
where
\beq
\text{I}_1= 2\pi ^2r(h-z)R^2\, ,
\eeq
\beq
\begin{split}
\text{I}_2=-2\int_0 ^1 4\pi r & (h-z)(R-ux)\\
&\times\arccos\biggl[\frac{2xR-u(x^2-1)}{2(R-u)}\biggr]dx,
\end{split}
\eeq
and $u=\sqrt{r^2-z^2}$.
$F(r)$ is presented in the inset of Fig.~\ref{fig:RDF}.
$R$ is here the length of $\bR$, i.e., the radial distance of a grain 
center from the cylinder axis.
The width of the spherical shells were chosen so that the standard error
of distribution from the mean was less than two per cent of the RDF, 
for the second peak. This gave the following resolutions of the RDFs; 
 $dr/d=0.1$ (ductile ensemble) and $dr/d=0.02$ (Finney). 

\subsubsection{\label{sec:coord3d} Coordination numbers}
Coordination numbers in three dimensions were found experimentally for 
the ductile ensembles, as described in section \ref{sec:exp3d}. 
As the grains in contact to the walls were known, the distributions of 
coordination numbers for the ductile ensembles are based only on the 
internal grains. 
In the second of the large ensembles ($c=0.75$), only grains at a minimum 
distance of three grain diameters from any boundary were included. 
For the Rampage and Finney ensembles, the coordination numbers were 
based on the position data. 
In the Rampage ensembles grains in a distance less than one grain diameter
apart were considered touching, as the model allows small overlaps 
between grains.
In the analysis of the Finney ensemble, grains were considered touching if 
the center to center distance $r$ was less than 1.02 grain diameters. 
The choice of this distance was based on the average contact number obtained
when different $r$'s were assumed for touching neighbors:
If grains were assumed touching only when their center to center 
distance was 1 $d$ or less, $\langle k \rangle =0.9$.
$\langle k\rangle$ rapidly increased as the assumed distance for touching
grains increased, and at 1.02 $d$, the average coordination number was 6.72.
Also, at 1.02 $d$, all grains (except for one) had at least three 
neighboring grains
at distances closer or equal to $r/d=1.02$, which is the stability 
criteria in 3D for grains shielded by granular bridges.

\subsection{\label{sec:results3d}Results (3D)}
\begin{figure}
\epsfig{file=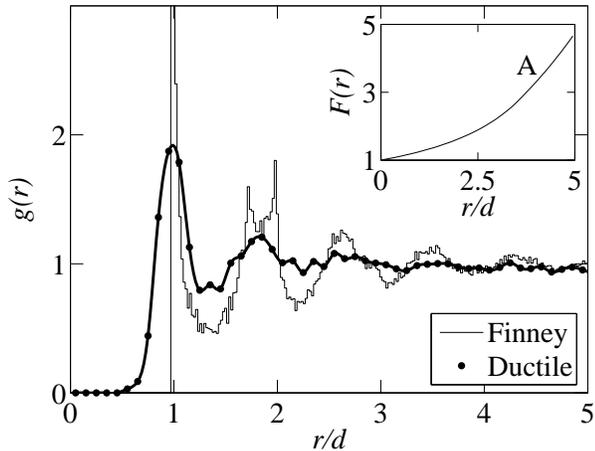,width=8.6cm}
\caption{The radial distribution functions for an ensemble of 
ductile grains ($\bullet$, bold line) and for the Finney ensemble of dry, 
hard grains (thin line). The bold line was obtained by splines to the data.
The inset shows the normalization function $F(r)$, see Eq. (\ref{eq:F(r)PD}), 
(curve A) used to correct 
for the 3D finite size effects for the ductile system, in the region 
$r/d\in[0,5]$.}
\label{fig:RDF}
\end{figure}

To calculate RDFs, the center positions of the grains must be known, whereas 
in the 3D ductile experiments the positions of the top points of 
the grains were measured.
Although the top points are not the shifted positions of the grain 
centers, they were used in calculating the RDF, as they represent 
correlations between  specific points in the ensemble in the same way that 
the center positions do.
Also, it should be directly comparable to the RDFs of hard granular ensembles, 
as in these the top positions are truly linear translations of the center 
positions. 
The detailed structure of an ensemble can only appear in the RDF if
there is a sufficient amount of grains in the ensemble. 
Only the large ductile system contains enough grains to 
capture the main features of the RDF, thus unfortunately, we can only 
compare the ductile system at a packing fraction of $c=0.70$ to the 
Finney ensemble ($c=0.636$), and only at a resolution 
of $dr=0.1\,d$. 
The normalized RDF of the ductile ensemble is shown in Fig. \ref{fig:RDF}, as 
is also the RDF of Finney's ensemble for dry granular media. 
From the figure, we see that the ductile ensemble is much less ordered than
the hard granular ensemble, as all 
peaks and valleys in the RDF are small compared to the Finney RDF.
We also note that the RDF of the ductile packing has a value for distances 
smaller than one grain diameter, which is not surprising, as this reflects the 
grains' ability to deform and thus obtain distances closer than one grain 
diameter.
We see fewer clear peaks for the ductile ensemble than for the Finney 
ensemble, as deformed grains broaden the peaks.
The split second peak in the RDF of the Finney ensemble
is not present in the ductile ensemble.

\begin{figure}
\epsfig{file=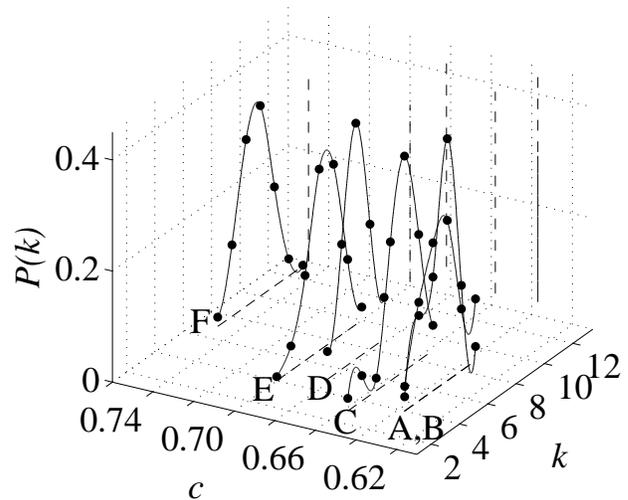,width=8.6cm}
\caption{The distribution of coordination numbers for the different 
ductile ensembles, as function of packing fraction. The curves A--F
are splines to the data points of the packings at packing fractions
$c=$0.64, 0.64, 0.66, 0.68, 0.70, and 0.75, respectively.}
\label{fig:Cdev}
\end{figure}

\begin{figure}
\epsfig{file=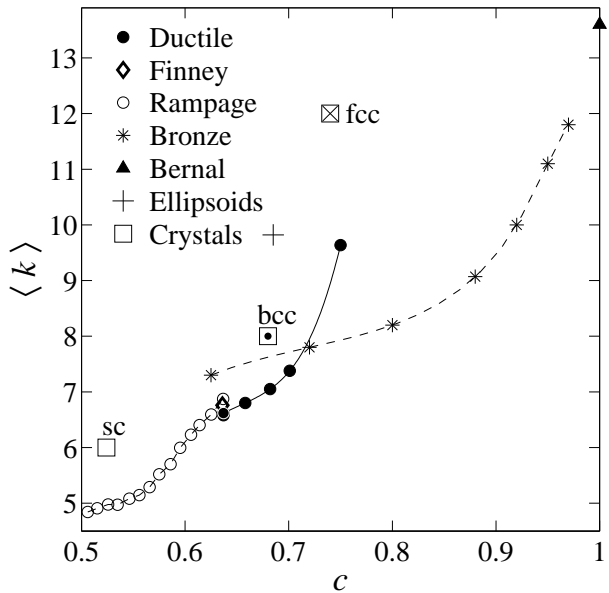,width=8.6cm}
\caption{Mean coordination numbers $\langle k\rangle$ as a function of packing 
fraction $c$ for several kinds of packings: The ductile ensembles ($\bullet$),
the Finney ensemble ($\diamond$), Rampage ensembles ($\circ$),
compacting bronze powder \cite{pap:Fischmeister78} ($\ast$), 
compacted plasticine \cite{InColl:Bernal65} ($\blacktriangle$), hard ellipsoid 
ensemble \cite{pap:Donev04} ($+$). 
The data for some crystalline structures of hard spheres are shown 
for reference: Simple cubic ($\square$), Body centered 
cubic ($\boxdot$), and Face centered cubic ($\boxtimes$).}
\label{fig:cvsC}
\end{figure}

The coordination number distributions for the ductile systems are presented in 
Fig. \ref{fig:Cdev}.
The distributions shift to higher coordination numbers as the density 
increases, as is to be expected. 
One interesting observation is that two of the distributions for the 
small ensembles contain grains of a coordination number as low as three.
This is possible for grains that are shielded by granular bridges, 
which are also found in the Finney and Rampage packings to a similar 
degree ($1.6\%$ of the internal grains). 
Due to the cohesion between 
the ductile grains, the effect of granular bridging was expected to be
more dominant in the looser ductile structures than in the Finney and 
Rampage structures, which is not the case.
For the denser ductile structures ($c>0.64$) all grains have
coordination numbers larger than three, but the width of the 
distributions remain the same within 4\%.
Figure \ref{fig:cvsC} shows the mean coordination numbers as a 
function of packing fraction for the ensembles presented in this 
paper, and also for a few cases from the literature.
The compaction of the ductile grain ensembles (filled circles) 
start in the neighborhood of the random close packed Finney 
ensemble (diamond), and then $\langle k\rangle$ increases with $c$.
The first five points on this curve were obtained from gravitationally
compacted ensembles, thus their compaction process is the same.
The last point, at $c=0.75$, was obtained by mechanical compaction of
a piston, and therefore does not necessarily represent the natural
evolution of the gravitational ensembles.
The numerical model (Rampage) evolved along the curve represented by
open circles in Fig. \ref{fig:cvsC}, during one simulation.
It ends up also in the neighborhood of the Finney ensemble. 
Note that the trajectory of the Rampage packing does not seem to be 
continued by the ductile ensembles, as the curves have different 
slopes at $c=0.64$.
Another ductile ensemble from the literature is presented in the 
figure (asterisks): 
Fischmeister et. al \cite{pap:Fischmeister78} conducted compaction experiments
on bronze powder by die compression at various pressures.
The compacted ensemble was forced open after compression, and the number 
of contacts and porosity measured.
For the smallest packing fraction, the powder was sintered in order to 
mark the contact points of a non-compressed ensemble.
The data at higher packing fractions were compacted at pressures 
of 0.2, 0.4, 0.6, 0.8 and 1.0 GPa, respectively.
The powder had an initial size distribution of $250$--$315\, \mu$m.
The packing of hard ellipsoids \cite{pap:Donev04} is marked as 
a plus sign in Fig. \ref{fig:cvsC}, and the crystalline ensembles of 
simple cubic, body centered cubic and face centered cubic are plotted
as various types of squares. 

\section{\label{sec:discussion} Discussion}
That the shape of the grains is an important parameter for the structure
is evident from the experiment by Donev et al. \cite{pap:Donev04}, who
studied the coordinations in an ensemble of ellipsoids and found a
mean coordination number of 9.82 in an ensemble of packing fraction 
$c=0.685$.
This value is much larger than that found in spherical ensembles 
at similar (but lower) $c$.
The compaction of hard spherical grains to higher densities than the 
random close packed (rcp) value ($c=0.63$--0.64) 
\cite{pap:Scott62,pap:BernalMason60} 
occurs by increasing the amount of crystalline regions.
Pouliquen et al. \cite{pap:Pouliquen97} obtained a strongly crystalline 
packing ($c=0.67$) from mono-disperse glass spheres by horizontal 
shaking and slow filling of a container.
However, handling hard spheres without fine-tuning the assembling
procedure commonly results in values around $c=0.64$.
The compaction of deformable grains allows larger packing fractions
to be obtained than hard grain ensembles do, because the shape of 
individual grains can change.
Whereas the crystallization of compacting hard grains would show up 
in the RDF, the evolution of the RDF during compaction is not given in 
ductile ensembles, as they might compact by structural ordering
or grain deformation.
Shape changes from that of a sphere or a disk are expected to cause 
disordering (e.g., there are no crystalline arrangement of pentagons).
There are space filling crystalline structures (squares, pyramids), 
but these are very unstable under any geometrical perturbation, as
more than three edges meet at each vertex of the structure.

The radial distribution functions in two and three dimensions
as they were studied here, were both seen to lose some of the 
structure associated with hard grain arrangements as high packing 
fractions were approached.
The RDF in 2D numerical hard disk ensembles has been shown
 \cite{pap:Myroshnychenko05} to develop its structure considerably 
through packing fractions $c=0.3$--0.83. 
The split second, the third, and the fourth peaks were seen to develop 
between $c=0.7$--0.83, and the RDF at $c=0.83$ resembles the
RDF of the 2D ductile experiment in the initial stage 
(Fig. \ref{fig:RDF2d}). 
Our 2D ductile system, when further compacted, gives less distinct 
splitting of peaks, and the RDF thus loses some of its detailed structure
toward $c=0.99$ (see Fig. \ref{fig:RDF2d}).
Interestingly, we saw that the peak at $r/d=1.695$ corresponded to 
a cluster of four grains in the configuration shown in Fig. \ref{fig:RDF2d}
in the initial part of the experiment, the same configuration that 
contributes to the peak at $r/d=\sqrt 3$ in hard disk ensembles.
Thus, the initial part of our experiment is dominated by similar 
configurations as in hard disk ensembles, but with shortened distances.
At the end of our experiment, though, several different configurations
contribute to this peak in the RDF, thus there are important structural
changes compared to hard disk packings.

A similar evolution is expected for 3D ensembles; 
A certain packing fraction must be reached for locally ordered grain 
arrangements to occur, and for details in the RDF structure to emerge.
When grain deformation becomes important for further compaction, 
the RDF structure becomes less distinct (Fig. \ref{fig:RDF}).
This `destructuring´ should become influential in the 3D ductile grain
ensembles at packing fractions close to the random close packing, 
$c\simeq0.64$, assuming that frictional forces between the grains 
prevent rearrangements.
The RDF of the ductile 3D system at $c=0.70$ has no clear peaks after
$r/d=3$, and has much smaller and broader peaks than that of the 
2D system at $c=0.99$.
This difference could be due to the extra degree of freedom introduced
by the third dimension, or possibly the differences in rheology between
the materials used in 2D and 3D, and how they react to the way they
were compacted.
Also a size distribution of the grains would tend to broaden the peaks
in the RDF, but the size distribution is similar for our 2D and 3D
experiments (3\% and 2.3\%, respectively).

The evolution of the coordination number distribution suggests
geometrical ordering to take place in the 2D ductile system.
The distribution narrows considerably, and only the number of 
grains with six contacts increases, see Fig. \ref{fig:Cdev2d}.
Grains with only two contacts occur for a limited range of 
packing fractions $c=0.906$--0.918, and these are due to 
shear motion, during which two neighboring ordered regions 
align.
In three dimensions, the evolution of the coordination number distribution
is not as distinct.
The average, $\langle k\rangle$, increases, but the width of the 
distribution stays practically constant during compaction.
The reason for this could be that the 3D ductile systems are
not as compacted as the 2D system is, thus a marked narrowing of 
the coordination number distribution might not take place until 
larger packing fractions are reached.
Also, the mean coordination number in 2D has a maximum
of six, while no established maximum exists in 3D.
Thus, the distribution of $k$ in 2D must narrow as the system gets
denser, whereas the 3D system does not have this strict constraint.

In 2D, the evolution of the average coordination number $\langle k\rangle$ 
increased with packing fraction $c$ toward the theoretical value
$\langle k\rangle=6$, which applies to 2D space filling structures 
(Voronoi cells).
At the end of the 2D experiment, 1\% of pore space remained, and 
according to Fig. \ref{fig:cvsC2d}, the increase in average coordination number
must be steep for $c$ in the range 0.99--1.
The evolution of $\langle k\rangle$ follows the straight line
between the square lattice ($\langle k\rangle=4$) and the space filling
structure ($\langle k\rangle=6$) up to $c=0.97$.
Although an increase of $\langle k\rangle$ with $c$ was expected, the linear
evolution was not.
The dominant crystalline structure is triangular from the start of the 
experiment, and regions of dense triangular structures ($k=6$) were
thought to develop rapidly during the initial compaction.  

3D deformable grains will approach a space filling structure of 
polyhedra with increasing packing fraction.
This structure is not necessarily the packing's Voronoi structure,
as the interfaces between neighboring grains then would 
have to be normal to the center-to-center distance.
However, the statistical features of the Voronoi structure might 
be applicable at $c=1$.
No theoretical value exists for the average coordination at $c=1$ 
in 3D, but a value around 13.6 
was found by Bernal \cite{InColl:Bernal65} based on an experiment 
with compacted plasticine.
The Voronoi tessellation of hard sphere packings at different packing 
fractions have shown that the average number of sides of a Voronoi cell, 
$\langle f\rangle$, decreases with increasing packing fraction
\cite{pap:Yang02,pap:Oger96}.
From Voronoi analysis of the ductile ensemble at $c=0.70$, we find 
that $\langle f\rangle=14.5$, and presume that $\langle f\rangle$ in the 
ductile ensembles at $c=1$ must be below this value.
At a packing fraction of $c=1$, all the faces of the polygon must be
touching a neighbor, thus $\langle f\rangle =\langle k\rangle$ at $c=1$.
This suggests that had the ductile ensembles been allowed to compact
to $c=1$, their average coordination number would be less than 14.5,
and possibly close to Bernal's experimental value of 13.6.

How the Voronoi cell structure changes during compaction must  
depend on the compaction procedure.
If the packing is compacted isomorphically, the relative positions of 
the grains are fixed with respect to each other, and the Voronoi 
structure is constant.
For grains on the square lattice, the Voronoi cells would remain 
squares throughout an isomorphic compaction, thus $\langle k\rangle=4$
for all $c$.
A random initial packing, or a non-isomorphic compaction procedure, 
would instead result in an increase of the average coordination number
toward $\langle f\rangle$.
Thus the trajectories a compacting system makes in the 
($\langle k\rangle ,\,c$)-plane is not fixed, but is
expected to depend on initial configuration and compaction procedure.
As seen from Fig. \ref{fig:cvsC},
the ductile Play-Doh ensemble compacts in a self-compacting
trajectory (apart from the last point on the curve), which evolves
differently than the compacted bronze powder.
The bronze powder was compacted by increasing uniaxial pressure, 
while our system sustained constant uniaxial pressure of gravity.
Also, the Rampage model of compaction has a trajectory leading into 
the region where the ductile compaction starts, but has a different 
slope at $c=0.64$ than the trajectory of the ductile packings.
The different compaction procedures might be an explanation for these 
different trajectories in the ($\langle k\rangle ,\,c$)-plane.

Hard spherical ensembles can rather easily be compacted to 
the rcp value, $c\simeq0.64$, and are well represented by Finney's 
packing in the ($\langle k\rangle ,\,c$)-plane.
This point might be fairly common for spherical ensembles, 
considering that both the Rampage model, and two of the ductile 
ensembles plot in the immediate neighborhood.
Looser initial configurations might be obtained in packings of high
cohesion or friction, which would change the starting
point of the compaction in the ($\langle k\rangle ,\, c$)-plane.
That the bronze powder of Ref. \cite{pap:Fischmeister78} plots 
at a higher $\langle k\rangle$ than the rcp is thought to be due to the 
wide size distribution of grains, and the sintering procedure used.
The hard elliptical grain ensemble \cite{pap:Donev04} was assembled in 
the same way as the Finney packing, but produced a much higher 
packing fraction and average coordination number.
The ability of the grains in the ductile packing to 
change their shape does not, however, take the ductile packing into the 
ellipsoid value of $\langle k\rangle=9.8$  until $c>0.75$ for the 
Play-Doh ensembles, and $c\sim 0.9$ for the bronze powder (see Fig.
\ref{fig:cvsC}).
Hence, the initial grain shape is an important factor for 
the initial compaction state in ($\langle k\rangle ,\, c$)-plane.

\section{\label{sec:conclusion} Conclusion}
We have performed compaction experiments on two- and three-dimensional
packings of mono-disperse ductile grains, and studied the evolution
of their structure with increasing packing fraction.
The radial distribution function and coordination number distribution
were found for both systems.
The radial distribution function developed broader and smoother peaks,
seemingly loosing ordered structure in both 2D and 3D.
The 2D local configuration around grains in distances $r/d=1.695\pm 0.05$
revealed that while mainly one configuration contributed to the 
corresponding peak in the initial RDF, two configurations 
contributed to the same peak at the most compacted stage of the 
experiment.
The coordination number distribution narrows considerably in the 
2D-packing during the compaction, while no such narrowing is seen
in the 3D-system.
The average coordination number in 3D evolves beyond that of 
hard spherical ensembles, and its evolution is discussed in 
relation to other granular ensembles, as the compaction procedure and
initial grain shape seems to be important parameters for the 
compaction trajectory in the ($\langle k\rangle ,\, c$)-plane.

This is a new approach to the study of compaction in granular ensembles,
which emphasizes the importance of grain ductility on the evolving
structure.
Further work is necessary to understand the importance of compaction 
procedure, initial grain shape, and rheology for the final structure.
In particular, 3D in-situ experimental investigations
of evolving geometry during compaction would give a strong contribution 
to the understanding of how local structure depends on compaction procedure.

%
%
 
\acknowledgments The project has been supported by the Norwegian
Research Council through the {\em Fluid Rock Interaction} {S}trategic
{U}niversity {P}rogram (grant 113354-420). LU would like to thank 
Birger Sevaldson for use of the MicroScribe.

\end{document}